\documentstyle[12pt]{article}

\begin{document}

\author{M. Apostol  \\ %EndAName
Department of Theoretical Physics,\\Institute of Atomic Physics,
Magurele-Bucharest MG-6\\POBox MG-35, Romania\\e-mail: apoma@theor1.ifa.ro}
\title{On defects in solids }
\date{J. Theor. Phys. {\bf 11} (1996) }
\maketitle

\begin{abstract}
The thermodynamical potential for dilute solutions is rederived, generalized
and applied to defects in solids. It is shown that there are always defects
in solids, {\it i.e.} there is no perfect solid at any finite temperature.
Apart from the temperature-dependent concentration of defects, another case
is presented, where the defect concentration does not depend on temperature.
\end{abstract}

With standard notations we may write 
\begin{equation}
\label{one}dE=-pdV+TdS+\mu dN 
\end{equation}
at equilibrium, whence 
\begin{equation}
\label{two}d(E+pV-TS-\mu N)=Vdp-SdT-Nd\mu \;\;. 
\end{equation}
We may introduce the thermodynamical potential 
\begin{equation}
\label{three}\Theta =E+pV-TS-\mu N\;\;\;, 
\end{equation}
and, noticing that $E,\;V,\;S,\;N$ are extensive, while $p,\;T\;$and $\mu $
are intensive quantities, we have $\Theta =\partial \Theta /\partial N\cdot
N $\ . On the other hand, we have from $\left( 2\right) $ $\partial \Theta
/\partial N=0$ at equilibrium, whence $\Theta =0$; which corresponds to the
definitions $\Phi =E+pV-TS=\mu N$ of the thermodynamical potential $\Phi $ .

\medskip\ 

Suppose that in a solid of $N_0$ sites we are interested in the existence of 
$n$ defects at fixed $p,T$ and $\mu $, where $n\ll N_0$, but both $n$ and $%
N_0$ are very large. These defects are supposed to be structureless, and may
be exemplified by vacancies, broken bonds, substitutional impurities, etc.
(In other cases, as for interstitial atoms for instance, we may use the
potential $\Phi $, corresponding to fixed $N$, with the same results as
those to be derived below.) For each of them we have a variation $%
\varepsilon $ of the energy, a variation $v$ of the volume and a heat $q$,
such that 
\begin{equation}
\label{four}\varepsilon =-pv+q+\mu \;\;. 
\end{equation}
For instance, when a vacancy is created, for which $v$ and $\mu $ appear in $%
\left( 4\right) $ with the minus sign, the solid receives the heat $q$ and
the mechanical work $-pv$ (the latter by the relaxation of the atomic
surrounding), increases its energy by $\varepsilon $ and loses an amount of
energy equal to the chemical potential $\mu .$ On the other hand, there is
an increase in entropy 
\begin{equation}
\label{five}S=\ln C_{N_0}^n\;\;\;, 
\end{equation}
where $C_{N_0}^n=N_0!/n!\left( N_0-n\right) !$, due to the disorder created
by the $n\;(\ll N_0)$ defects. For $n$ and $N_0$ large enough and $n\ll N_0$
we may also write 
\begin{equation}
\label{six}
\begin{array}{c}
C_{N_0}^n= 
\frac{N_0!}{n!(N_0-n)!}=\frac{N_0^{N_0}}{n^n(N_0-n)^{N_0-n}}=\left( \frac{N_0%
}n\right) ^n\cdot \frac{\left( 1-n/N_0\right) ^n}{\left( 1-n/N_0\right)
^{N_0}}\approx \\ \approx \left( \frac{N_0e}n\right) ^n\;\;\;, 
\end{array}
\end{equation}
whence 
\begin{equation}
\label{seven}S\approx n\ln \left( \frac{N_0e}n\right) \;\;. 
\end{equation}
The thermodynamical potential $\Theta $ given by $\left( 3\right) $ may
therefore be expressed as 
\begin{equation}
\label{eight}\Theta =\Theta _0+(\varepsilon +pv-\mu )n+Tn\ln \left( \frac
n{N_0e}\right) \;\;, 
\end{equation}
where $\Theta _0$ is the value of $\Theta $ at $n=0$ (homogeneous phase);
or, using $\left( 4\right) $, we have 
\begin{equation}
\label{nine}\Theta =\Theta _0+qn+Tn\ln \left( \frac n{N_0e}\right) \;\;. 
\end{equation}
This is the generalization for the potential $\Theta $ of the well-known
expression\cite{Lan} of the thermodynamical potential $\Phi $ ({\it i.e}.at
fixed $N$) for dilute solutions.

\medskip\ 

At equilibrium we should have $\Theta =0$ and $\partial \Theta /\partial n=0$
. Here we distinguish two cases, but first we remark that the heat $q$
should always be positive; otherwise the solid would be unstable. Then, we
remark that an entropy $s$ is created by producing a defect, accounted by
the degree of local disorder generated on this occasion (which amounts to
saying that actually the defects have a structure). If the process of
creating the defects is a non-equilibrium process, {\it i.e.} if the defects
are created fast enough in comparison with the relaxation time of the solid,
then $q<sT$ and , from the equilibrium condition, we have 
\begin{equation}
\label{ten}n=N_0e^{-q/T} 
\end{equation}
and $\Theta _0=Tn$; which means that we always have defects in solid (fot
any $T\neq 0$), whose concentration increases with increasing the
temperature, as it is well-known. In the second case, the defects may be
created by an equilibrium process, {\it i.e.} by a process that is slow
enough in comparison with the relaxation time of the solid, so that $q=sT$ .
From the equilibrium conditions we get in this case 
\begin{equation}
\label{eleven}n=N_0e^{-s}\;\;\;, 
\end{equation}
{\it i.e.} there always exists (for any $T\neq 0$) a finite concentration of
defects, which is independent of temperature. In general, for any realistic
values of $q$ and $s$, and normal temperatures, the defect concentration
given by both $\left( 10\right) $ and $\left( 11\right) $ is extremely low.

\medskip\ 

The first case, given by $\left( 10\right) $, of defect concentration
increasing with temperature, represents a rather common situation, which is
well documented. The second case, of a constant defect concentration as
given by $\left( 11\right) $, has been brought into discussion more
recently. For example, a constant, temperature independent concentration of
vacancies has been invoked \cite{Apo1}$^{,}$\cite{Apo2} in some alkali doped
fulerides, in order to explain a certain feature in the $NMR$ spectrum of
the alkali nuclei, called the $T-T^{^{\prime }}$ splitting.

\medskip\ 

Perhaps it is also appropriate to discuss here another aspect of this
problem, concerning the relevance the defects may bear on the existence of
the low-dimensional solids. The results presented here show that defects are
always present in solids at any non-vanishing temperature. For a
one-dimensional solid this means that the solid may be fragmented into
pieces, of a certain average length, as given by $\left( 10\right) $ or $%
\left( 11\right) $ . However, it may happen, as for instance in the first
case $\left( 10\right) $ at low enough temperatures, that this average
length be still large enough for the fragments have a thermodynamical
behaviour. In this case it is known that a one-dimensional solid,
irrespective of temperature, does not exist, in the sense that it is
unstable with respect to the propagation of the long-wavelength phonons. A
similar situation appears for a two-dimensional solid at $T\neq 0$, though
here the defects are irrelevant (as well as in three dimensions) for the
propagation of the long-wavelength phonons. It has been shown that the
realistic experimental conditions always require a certain constraint upon
the low-dimensional solids, which render them stable under the propagation
of the long-wavelength phonons, so that the question of the melting of these
solids is meaningful. All these questions are discussed in Ref.4,5, where
the melting temperature has been estimated by means of a mean-field theory
as being that temperature beyond which the solid gets soft and is no longer
able to bear phonons. Very likely this melting temperature is an
overestimation, and the melting defined in this way is intertwined with a
continuous transition toward a state with a variable range of crystallinity
in one dimension (as indicated by $\left( 10\right) $ for example), and with
a subsequent state with extended, topological defects in two dimensions.

\end{document}